\documentclass{article}

\usepackage{arxiv}

\usepackage[utf8]{inputenc} 
\usepackage[T1]{fontenc}    
\usepackage{hyperref}       
\usepackage{url}            
\usepackage{booktabs}       
\usepackage{amsfonts}       
\usepackage{nicefrac}       
\usepackage{microtype}      
\usepackage{lipsum}
\usepackage{amssymb}
\usepackage{amsmath}
\usepackage{graphicx}
\usepackage{caption}

\title{\textbf{Data-driven Contact Network Models of COVID-19 Reveal Trade-offs between Costs and Infections for Optimal Local Containment Policies}}

\author{
    \sffamily\large \vspace{0.15in}
    Chao Fan\textsuperscript{1,*}, Xiangqi Jiang\textsuperscript{2}, Ronald Lee\textsuperscript{2}, Ali Mostafavi \textsuperscript{1,*}\\
    \sffamily\normalsize
    \textsuperscript{1}Department of Civil and Environmental Engineering, Texas A\&M University, College Station, TX, 77843, U.S.\\
    \sffamily\normalsize
    \textsuperscript{2}Department of Computer Science and Engineering, Texas A\&M University, College Station, TX, 77843, U.S.\\
    \sffamily\normalsize
    \textsuperscript{*}Corresponding authors: chfan@tamu.edu, amostafavi@civil.tamu.edu\\
}

\begin{document}
\maketitle

\begin{abstract}
\sffamily While several non-pharmacological measures have been implemented for a few months in an effort to slow the coronavirus disease (COVID-19) pandemic in the United States, the disease remains a danger in a number of counties as restrictions are lifted to revive the economy. Making a trade-off between economic recovery and infection control is a major challenge confronting many hard-hit counties. Understanding the transmission process and quantifying the costs of local policies are essential to the task of tackling this challenge. Here, we investigate the dynamic contact patterns of the populations from anonymized, geo-localized mobility data and census and demographic data to create data-driven, agent-based contact networks. We then simulate the epidemic spread with a time-varying contagion model in ten large metropolitan counties in the United States and evaluate a combination of mobility reduction, mask use, and reopening policies. We find that our model captures the spatial-temporal and heterogeneous case trajectory within various counties based on dynamic population behaviors. Our results show that a decision-making tool that considers both economic cost and infection outcomes of policies can be informative in making decisions of local containment strategies for optimal balancing of economic slowdown and virus spread. 
\end{abstract}


\vspace{0.3in}

\section*{\sffamily Introduction}
\vspace{-0.5em}
The coronavirus disease 2019 (COVID-19) has caused a global pandemic threatening public health and human well-being \cite{Headey2020}. As of November 4, 2020, 47.7 million confirmed cases and 1.2 million deaths have been reported worldwide \cite{JohnHopkinsUniversity2020}. In response to the pandemic and in the absence of effective vaccines and drugs, numerous non-pharmaceutical interventions, such as stay-at-home and social distancing orders, have been implemented in the majority of the affected countries and regions \cite{Kraemer2020,Chinazzi2020}. These measures were enacted to quarantine infected persons thus isolating cases and suppressing virus transmission \cite{Rader2020}. The execution of these measures, however, caused a significant economic slowdown \cite{Polyakova2020}. In particular, travel restrictions and business lockdowns have caused soaring unemployment and a decline in tax revenue \cite{AURAY2020104260}. Hence, it is essential to examine policy solutions for making an optimal trade-off between virus control and economic recovery, especially in hard-hit regions. 

As society faces this global public health challenge, numerous studies in a variety of research streams relevant to pandemic mitigation have been published. Departing from the cross-region spread of infection, one study \cite{Kraemer2020} investigated the effects of travel restrictions on delaying epidemic outbreaks and reducing the number of cases. This research was followed by a further study on understanding cross-region population flow as a driver of the spatial-temporal distribution of COVID-19 in China \cite{Jia2020}. Through the use of mathematical models on metapopulations, such as the global epidemic and mobility model (GLEAM) \cite{Balcan2009}, risk source model \cite{Jia2020}, and disease contagion models \cite{Newman2002,Chowell2004}, these studies explored the mechanisms affecting the large-scale pandemic and demonstrated the role of human behaviors (especially human mobility) in disease transmission, and also examined the effectiveness of non-pharmacological interventions in virus spread control. With the understanding of the underlying mechanisms influencing epidemic spread and ubiquitous population movement data, data-driven methods and tools, such as deep neural networks \cite{Ramchandani2020,kapoor2020examining}, have been adopted to predict the spread of infections across states and counties. The well-performed predictions obtained through the use of these methods expand the capability of forecasting future trajectories of increases in the number of cases across different regions \cite{Chang2020}. 

Despite enormous progress \cite{Holtz2020} in slowing down the cross-regional spread of the pandemic, the COVID-19 pandemic has already entered the phase of community spread. The continuing infection growth accentuates the need for further localized containment measures \cite{schlosser2020covid19}. Existing meta-population-based analyses, however, are limited in capturing high-resolution person-to-person transmission to inform the evaluation of localized measures \cite{Balcan2009}. Recent studies have argued that anonymized mobile phone data, when used properly and carefully, could characterize the epidemic dynamics among human contacts during all stages of the pandemic life cycle \cite{Oliver2020}. Mobility data collected from a large number of devices enables time-resolved characterization of population contact patterns, making it possible to probe the mechanisms by which disease is transmitted among the population with precision unattainable by other data sources, such as surveys. Due to the benefits of mobile phone data \cite{Grantz2020}, studies \cite{Liu2018a} have attempted to simulate the SARS-CoV-2 transmission on synthetic populations derived from mobile phone data in normal situations and to model the impact of individual-based measures, such as testing, contact tracing and household quarantine \cite{Aleta2020a}. 

Human behaviors, however, are dynamic, as the pandemic evolves. Limited consideration of the dynamics of human contact patterns would undermine the robustness of localized measures of pandemic mitigation. Models that can explain the virus spread based on dynamic human behaviors are especially needed. In addition, after the execution of stay-at-home orders and business activity restriction policies in the United States during March and April 2020, many states and counties allowed reopening of economic activities to relieve the heavy burden on the economy caused by the shutdown \cite{Bonaccorsi2020}. Although lifting restrictions could help with economic recovery, the subsequent increase in population contact activities drove the pandemic to a new peak \cite{Kaxiras2020}. Clearly, an integrative consideration of infections and economic costs in the decision-making process is critical to inform making tradeoff decisions between the number of infections and the economic costs of containment policies. The existing models do not enable examining this important tradeoff.

In this study, we investigated the dynamic contact patterns from fine-grained, anonymized data for millions of mobile devices along with census and demographic data. Accordingly, we then synthesized weekly contact networks of the populations based on their dynamic contact patterns captured by mobile phone data from the date of first reported cases through the end of June 2020 in ten selected US metropolitan counties. Each agent in the contact networks is associated with a census block group (CBG) within specific residential areas. On top of the dynamic data-driven contact networks, we overlaid a time-varying, degree-based Susceptible-Exposed-Infectious-Recovered (SEIR) contagion model that simulates the case trajectories among the synthetic population. This model allows us to capture the temporal evolution of the pandemic and the spatial distribution of the infections due to the dynamic contact patterns of the population. Through the understanding of the spatial-temporal heterogeneity of disease spread, we could identify the most at-risk populations, quantify the cost and infection trajectories, and provide a quantitative analysis of the effects of combined local containment policies on the costs and infections. 

Here, we demonstrate the performance of our model in terms of out-of-sample prediction based on population contact behaviors. The results show multiple waves of virus spread and spatial variation within ten US metropolitan counties. The results also highlight the heterogeneous effects of local policies on contact networks and disease spread across different regions. With these important observations, we propose a method that allows us to quantify the cost of the local policies and recommends optimal combined policies to achieve a better trade-off between economic recovery and virus control.

\section*{\sffamily Results}
\vspace{-0.5em}
\subsection*{\sffamily Dynamic data-driven contact networks}
\vspace{-0.7em}
To provide a quantitative estimate for the dynamic contact patterns in the population of agents in each county, we first use detailed demographic data from US Census Bureau \cite{Bureau} to generate a certain number of agents and assign them to specific Census Block Groups (CBGs). During this process, we ensure that the numbers of agents in the CBGs are proportional to the numbers of residents according to the census data. Virus transmission occurs through the co-presence, defined as contact through activities, of two or more agents in the same small area. To simulate infection spread among the synthetic population, we build weekly contact networks ($\mathcal{G}$) that encode the dynamic contact patterns of anonymized mobile devices from our empirical data (see \textit{Methods}). The co-presence of two mobile devices in a point of interest (POIs) determines a link between them, and the duration of the co-presence is the weight of the link ($\omega_{ij}$, where $i$ and $j$ are different agents). POIs refer to public common places such as restaurants, museums, nature parks, day care services, gasoline stations. The synthetic contact networks follow the same patterns of degree distribution and duration of contact as what is documented in the empirical data on a weekly basis (Fig. 1).

\begin{figure}
  \centering
  \includegraphics[scale=0.5]{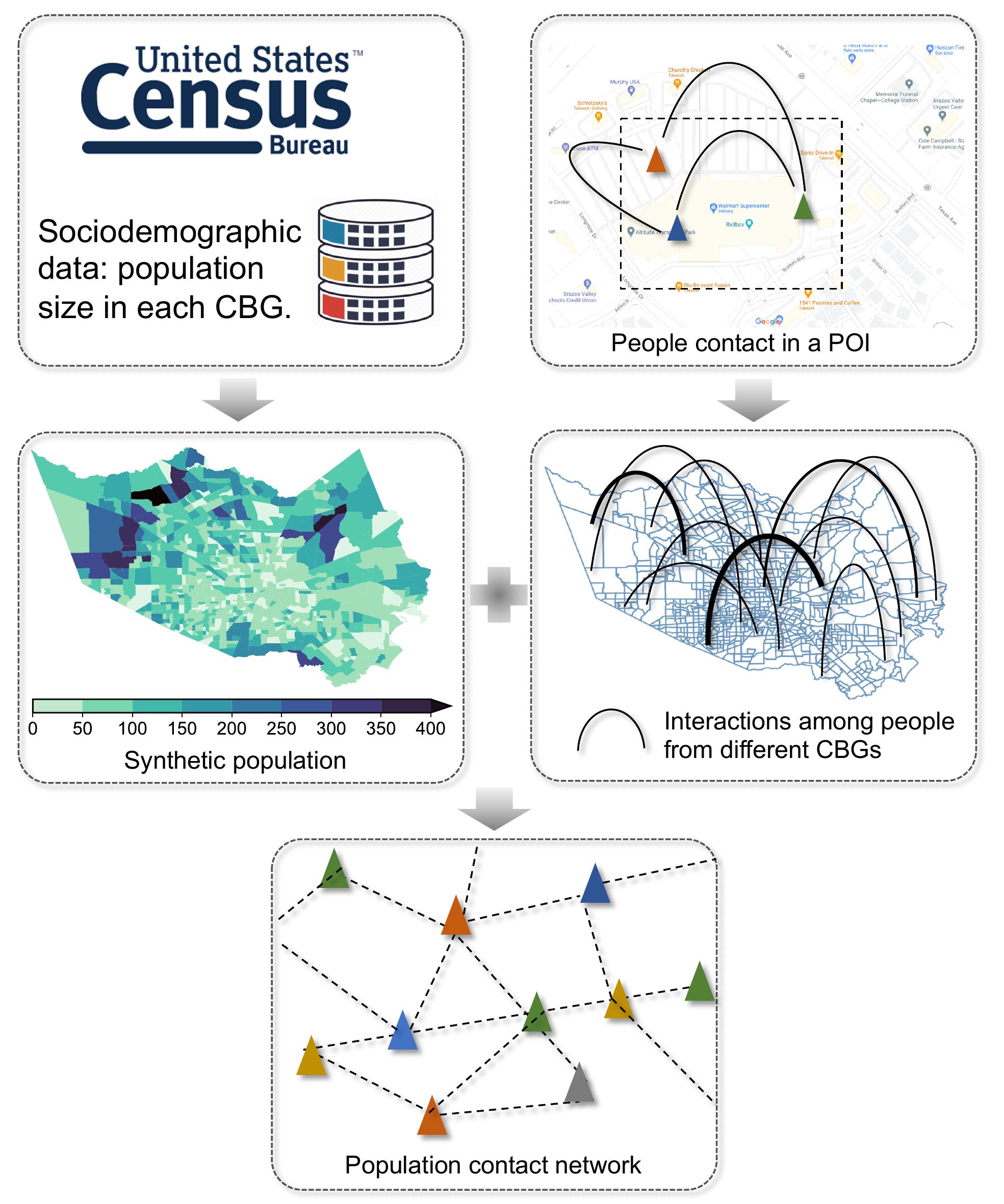}
  \caption{Schematic illustration of synthesizing the dynamic contact networks from anonymized mobile devices.}
  \label{fig:fig0}
\end{figure}

\subsection*{\sffamily Epidemic contagion in contact networks}
\vspace{-0.7em}
We implemented a stochastic time-varying and degree-based compartmental model in which the spread of infection relies on the structure of networks and the duration of contacts (Fig. 2a, the state transition diagram). The model is formulated as a system of ordinary differential equations. In this model, we consider that an agent could be in only one of the five states: susceptible ($S(t)$), exposed ($E(t)$), infectious asymptomatic ($I_a(t)$), infectious symptomatic ($I_s(t)$) and removed ($R(t)$) at any time-step $t$ (1d). Ostensibly, the susceptible agents are those who have not been infected at time $t$. We consider that the susceptible agents have a certain chance of getting infected based on their contact activities with exposed agents ($E(t)$) and infectious agents (asymptomatic infectious ($I_a(t)$) and symptomatic infectious ($I_s(t)$)) in the contact networks \cite{CentersforDiseaseControlandPrevention2020}. That is, the structures of the networks that facilitate the spread of the disease are incorporated, which leads to heterogeneous probabilities of susceptible agents to be infected \cite{PhysRevE.63.066117}. The agents with higher degrees (more links connecting to them) are more likely to be in contact with infections, and therefore they would be more likely to get infected \cite{barabasi2016network}. Assuming the agents with the same degree behave similarly \cite{barabasi2016network}, we employ degree block approximation to place the agents that have the same degree into the same block (group). This approach allows us to create a separate compartmental model in each group of agents based on their degrees. 

\begin{figure}
  \centering
  \includegraphics[width=17cm]{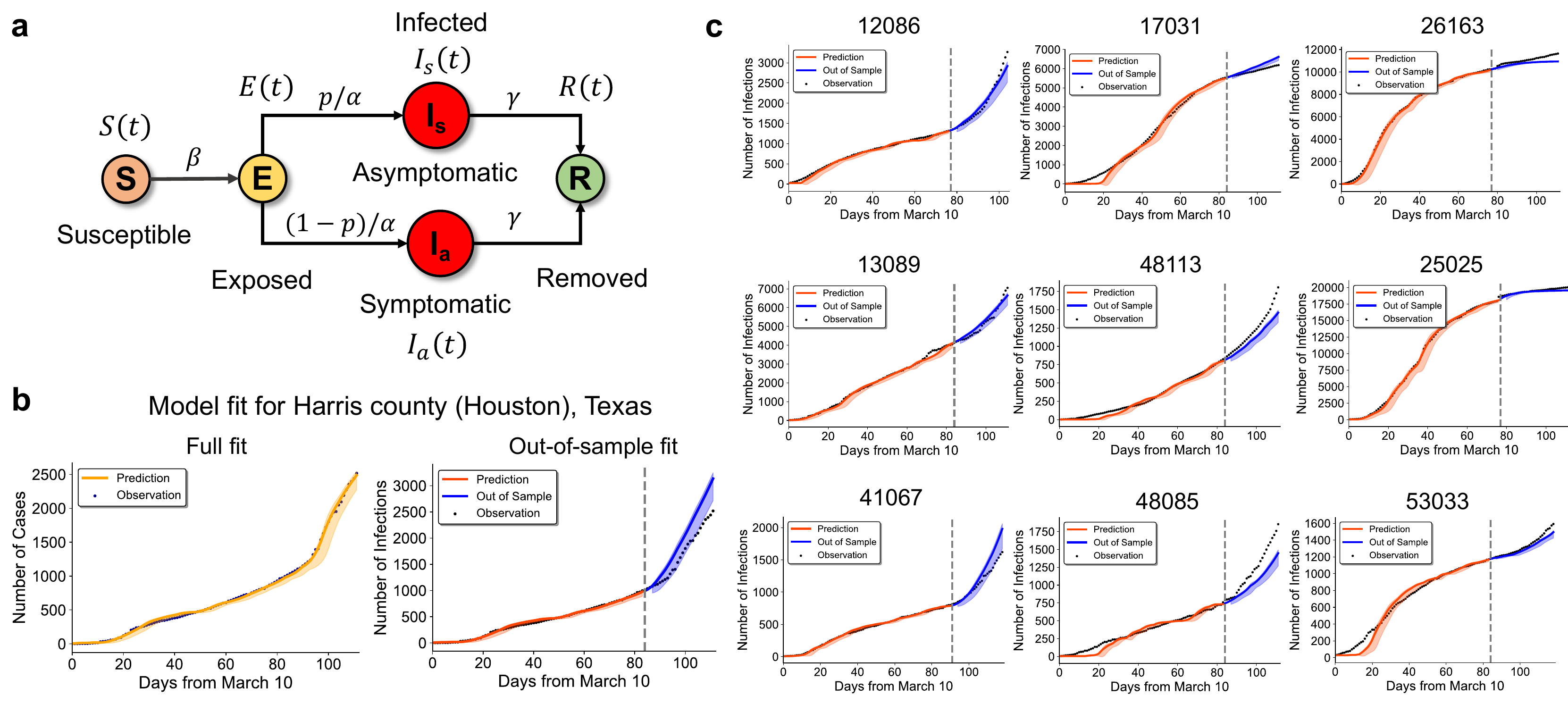}
  \caption{Model description and fit. {\sffamily \textbf{a.}} The state transition diagram illustrating the compartmental model used to capture the transmission process of the SARS-CoV-2 among the population in different states. Specifically, we consider the susceptible ($S$), exposed ($E$), infectious asymptomatic ($I_a$) and infectious symptomatic ($I_s$). The rates, $\beta$, $p/ \alpha$, $(1-p)/ \alpha$, and $\gamma$ indicate the probability of people transformed from the current state to the next state. Here, $p$ is equal to 20\%. More details for the model are provided in the main text and the supplementary information. {\sffamily \textbf{b.}} Model fit for Harris County (Houston area), Texas. The plot on the left panel is the full fit on the full range of data, while the plot on the right panel is the out-of-sample prediction. In the out-of-sample prediction, we calibrated the model on the data before 80 days since March 10 and predicted the infections afterwards. {\sffamily \textbf{c.}} The model fit and out-of-sample prediction for additional nine US metropolitan counties. The numbers on top of each subplot show the Federal Information Processing Standard (FIPS) codes of the selected counties.}
  \label{fig:fig1}
\end{figure}

Contact networks tend to lack degree correlations, which means the probability that a link points from an agent with degree $k$ to an agent with degree $k^{\prime}$ is independent of $k$. Such a heterogenous contact pattern in the group forms the basis to formulate a general differential equation system. The probability that a random chosen link connects an agent in the group of degree $k^{\prime}$ is the excess degree $k^{\prime} p_{k^{\prime}} / \langle k \rangle$, where $p_{k^{\prime}}$ is the probability of a random chosen agent with degree $k^{\prime}$ in the contact network. Existing infected agent should have at least one link connecting to another infected agent, the one that transmitted the disease. Therefore, the number of links available for future transmission of the agent is $(k^{\prime}-1)$. The fraction of infected nodes in the neighborhood of a susceptible agent in a group of degree $k$ is defined as a density function $\Theta_k$. We also consider the agents at both latent stage and the infectious stages can spread the disease and infect other susceptible agents. As such, the density function can be written as:

\begin{equation}
\Theta_k = \frac{\sum _{k^{\prime}} (k^{\prime}-1) p_{k^{\prime}}}{\langle k \rangle} (e_{k^{\prime}} + i_{k^{\prime}})
\end{equation}


where $e_{k^{\prime}}$ represents the density of exposed agents with degree $k^{\prime}$, $i_{k^{\prime}}$ represents the density of infected agents with degree $k^{\prime}$ in the network. Exposed agents are those who have been infected but are not yet infectious. As observed from equation (1), in the absence of degree correlations, the density function $\Theta_k$ is independent of $k$. By differentiating both sides of the equation, we can obtain:

\begin{equation}
\frac{d \Theta_k}{dt} = \sum _{k} \frac{(k-1)p_k}{\langle k \rangle} (\frac{d e_k}{d t} + \frac{d i_k}{d t})
\end{equation}

Exposed agents are transformed from susceptible agents into infectious at a rate of $1/ \alpha$. Hence, the change of the density of exposed agents depends on two transitions: the number of agents transitioned from susceptible; and the number of agents transitioned to the infectious. We formalize the changing rate of the density of exposed agents as:

\begin{equation}
\frac{d e_k (t)}{d t} = \beta k \Theta_k(t) s_k(t)-\frac{e_k (t)}{\alpha}
\end{equation}

where $s_k (t)$ represents the density of susceptible agents with degree $k$ in the contact network at time $t$, and $\alpha$ is an inverse of a rate, indicating the mean latent period for the disease. To be consistent with reported results in the literature \cite{Guan2020}, we consider that the SARS-CoV-2 has a latent period of 14 days in this study ($\alpha$ equals 14). The infection rate is proportional to the general infection rate $\beta$ in the entire contact network and the degree $k$, which specifies the chance of a susceptible agent with degree $k$ being infected.

Exposed agents will further move to infectious stage, leading to the change of density of infected agents in the network. The change of infected agents also depends on two transitions: one is agents transitioning from the exposed state, and the other is the agents transitioning to recovery status. Therefore, the changing rate of the density of infected agents at time t could be given as:

\begin{equation}
\frac{d i_k(t)}{d t}=\frac{e_k(t)}{\alpha} - \gamma i_k(t)
\end{equation}

where $\gamma$ is the recovery rate, meaning that both asymptomatic and symptomatic infections would be removed after a mean infectious period $1/ \gamma$. The infections with degree $k$ are composed of asymptomatic infectious and symptomatic infectious agents with degree $k$. Hence, the density of infections in the contact network $\mathcal{G}$ could be represented as:

\begin{equation}
i_k(t) = i_{ak}(t) + i_{sk}(t)
\end{equation}

where $i_{ak} (t)$ and $i_{sk} (t)$ represent the density of asymptomatic infectious and symptomatic infectious agents with degree $k$ in the network respectively. Based on the settings in an existing study \cite{Aleta2020}, we consider that 25\% of infections are asymptomatic infectious and the rest are symptomatic. This study is also aware of the case detection rate, which may vary with prevalence, testing capacity, testing protocols, and reporting fatigue. These factors may in turn all vary spatially and temporally. To take the influences of these factors into account, we calibrate the model in each week so that the model can capture the dynamic infectious situation well.

Then, with a probability of $p$, exposed agents in the latent state will move to the symptomatic infectious state; otherwise, they will move to asymptomatic infectious. Since the probability $p$ in our model is fixed, we still use $i_k (t)$ in the mathematical formulation of the problem. The probability $p$, however, influences the trajectory of disease spread in contact networks, which will be discussed later.

Plugging both equations (3) and (4) into equation (2), we have:

\begin{equation}
\frac{d \Theta_k (t)}{d t} = \sum _{k} \frac{(k-1) p_k}{\langle k \rangle} (\beta k \Theta_k(t) s_k(t) - \gamma i_k(t))
\end{equation}

Here, we keep only the first order terms, which means that the $s_k (t)$ could be ignored above, as for small t, $e_k (t)$ and $i_k (t)$ are much smaller than one and $s_k (t)$  is much close to 1. Then, the equation (6) could be simplified as:

\begin{equation}
\frac{d \Theta_k(t)}{d t}=(\beta \frac{\langle k^2 \rangle - \langle k \rangle}{\langle k \rangle} - \gamma) \Theta_k(t)
\end{equation}

To solve this equation, we get:

\begin{equation}
\Theta_k(t) = Ce^{t/\tau}
\end{equation}

\begin{equation}
\tau = \frac{\langle k \rangle}{\beta \langle k^2 \rangle - \langle k \rangle (\beta + \gamma)}
\end{equation}

where $\tau$ is the characteristic time for the model. Using the initial condition, $\Theta_k (t=0)=C$. Hence, 

\begin{equation}
C = \frac{\langle k \rangle -1}{\langle k \rangle} (e_0 + i_0)
\end{equation}

where $e_0$ and $i_0$ are initial values for the densities of exposed agents and infected agents.

The susceptible agents can only transform to exposed agents. Hence, we formulate the change of the density of susceptible agents as:

\begin{equation}
\frac{d s_k (t)}{d t} = \beta k \Theta_k (t) s_k (t)
\end{equation}

The removed agents include both agents recovered from the infectious state and agents who died of the disease. Therefore, we determine the changing rate of removed agents using:

\begin{equation}
\frac{d r_k(t)}{d t} = \gamma i_k(t)
\end{equation}

\begin{equation}
N_k / N = s_k(t) + e_k(t) + i_k(t) + r_k(t)
\end{equation}

where $r_k (t)$ is the density of removed agents with degree $k$ in the network, $N_k$ and $N$ are the number of agents with degree $k$ and the total number of agents in the contact network $\mathcal{G}$ respectively. The ordinary differential equation is formed in each group of agents with specific degrees. The total number of infected agents is the sum of all infected degree-$k$ agents: $I= \sum _{k} p_k i_k N$, where $p_k$ is the probability density of the agents with a degree of $k$. Hence, the equations above capture, with a set of equations for all degree-$k$ agents, the time-dependent behavior of the whole system. 

With the estimates of populations in each state, the next critical step is to determine which agent is in what state to capture both temporal and geographical patterns of the pandemic spread in the contact network. Here, we adopt the idea of network percolation in three transitions to simulate the state transition of specific agents. First, susceptible agents are more likely to get infected if they are exposed to (contact) infections for a long time. Hence, to select the agents who will make transition from susceptible state to exposed state in the next time step, we compute the time that susceptible agents spent in contact with infectious agents. The probability of a susceptible agent being infected $(p_i^{(S \to E)})$ is proportional to contact time with infectious agents. 

\begin{equation}
p_i^{(S \to E)} \propto (\sum _{j \in I} \omega_{ij})
\end{equation}

where $i$ is a susceptible agent and $j$ is one of the infectious agents.

Second, exposed agents in latent state have two directions for transitioning to the next state: symptomatic and asymptomatic infections. We assume that symptomatic infections tend to have fewer movements and contact activities, while asymptomatic infections may maintain their behaviors as normal. Based on these assumptions, we compute the differences in contact activities and how the exposed agents behave during the following week and the current week. The probability of an exposed agent having symptomatic infection is proportional to the differences in the contact activities. The more reduced the contact of an exposed agent in the following week, the higher the probability of transitioning to a symptomatic infectious state (rather than an asymptomatic infectious state). We denote the probability of transitioning to symptomatic infection $(p_i^{(E \to I)})$ as follows: 

\begin{equation}
p_i^{(E \to I_s)} \propto (\sum _j \omega_{ij}^{(T+1)} - \sum _j \omega_{ij}^{(T)})
\end{equation}

where, $\omega_{ij}^{(T+1)}$ is the contact time between $i$ and $j$ on week $T+1$. 

Third, both asymptomatic and symptomatic infectious agents would transition to removed agents. They may recover from the disease or die of it. The mean duration of infection is about 14 days, based on the information from the Centers for Disease Control and Prevention \cite{Healthcareworkers}. Hence, we consider the probability of an infectious agent being removed in the next time step follows a Gaussian distribution with a mean of 14 and a standard deviation of 3. 

\begin{equation}
p_i^{(I \to R)} \sim \mathcal{N}(14,3)
\end{equation}

Each agent in the state of infection would be assigned a probability at each time step, and the agents with a higher probability are more likely to be selected for state transition. 

Through the above-described contagion and percolation process, this model allows us to capture both temporal evolution and geographical distribution of the infections and how the epidemic spreads in human networks due to contact activities. To obtain model parameters over time, we calibrated the model for each week and in every county in the United States using cumulative confirmed cases obtained from the New York Times \cite{TheNewYorkTimes2020}. 

\subsection*{\sffamily Model fitting and validation}
\vspace{-0.7em}
To validate the model, we showed the performance of the model in fitting real case data and predicting out-of-sample cases using the model calibrated on the period before. Specifically, we split the study period into two parts: the first 12 weeks as the learning set for fitting and calibrating the model; the balance of the time (4 weeks) as the testing set for predicting the confirmed cases. We first fit the model taking the weekly contact networks in the first 12 weeks as the input and obtained the values of model parameters. Due to the delay in case testing and reporting, our model implemented on the contact network in any week was fitted on the cases in the following week by minimizing the root mean square error (RMSE). Hence, we define the loss function as:

\begin{align}
\begin{split}
\min_{\beta,\gamma} \mathcal{L}_{\mathcal{G}} = \sqrt{\sum _{d=1}^D \frac{(\hat{y}_d-y_d)^2}{D}} \\ s.t. (\hat{y}_d - y_d)^2 \leq 0.01y_d
\end{split}
\end{align}

where, $d$ is the day in the fitting process, $D$ is the number of days in total, $\hat{y}_d$ is the predicted number of cases, and $y_d$ is the actual reported number of cases. It should be noted that the constraints we added here were to ensure that the number of cases on the last day of the prediction was close to the actual reported cases. That is, the deviation of the predicted results would not influence the prediction of the following weeks. 

It is evident that using mathematical derivation to solve the problem of Eq. (17) and get analytical solutions is challenging. That is because the functions are usually not continuous and differentiable. A more efficient way is to calculate the numerical solutions through a heuristic algorithm. In this study, we employed the global pattern search algorithm as a derivative-free numerical optimization method to identify the optimal point which can minimize the loss and satisfy the constraint. The estimated values for model parameters and the optimized RMSE are shown in the supplementary information. 

Through the training process, we obtained the model parameters including $\beta$ and $\gamma$. The stochasticity is introduced in the model through the initialization of the exposed and infected agents as well as the specific infected agents in the groups of degrees. (See supplementary information for more details.) Running the resulting models on contact networks in the testing set, we predicted the confirmed cases in the testing weeks. Specifically, the parameters obtained from the most recent week in the training set for each county are used to make the predictions of out of sample data (testing set). We evaluated the predictive performance of the model by comparing the differences between the predicted cases and actual reported cases. Sub-figures in Figure 2b and 2c show that the model fits the out-of-sample case data very well, demonstrating that this model is effective in extrapolating beyond the training set to future periods. 

\subsection*{\sffamily Spatial-temporal heterogeneity of infections}
\vspace{-0.7em}
Since the synthetic population of the contact networks is generated based on US Census data together with the results of the model, we can characterize the differential spread of SARS-CoV-2 across the CBGs in a county. Through examining the spatial and temporal mechanisms of the disease spread, we are able to quantify different levels of infection in different places and provide a quantitative approach for modeling the execution of local containment and recovery policies. 

\begin{figure}
  \centering
  \includegraphics[width=17cm]{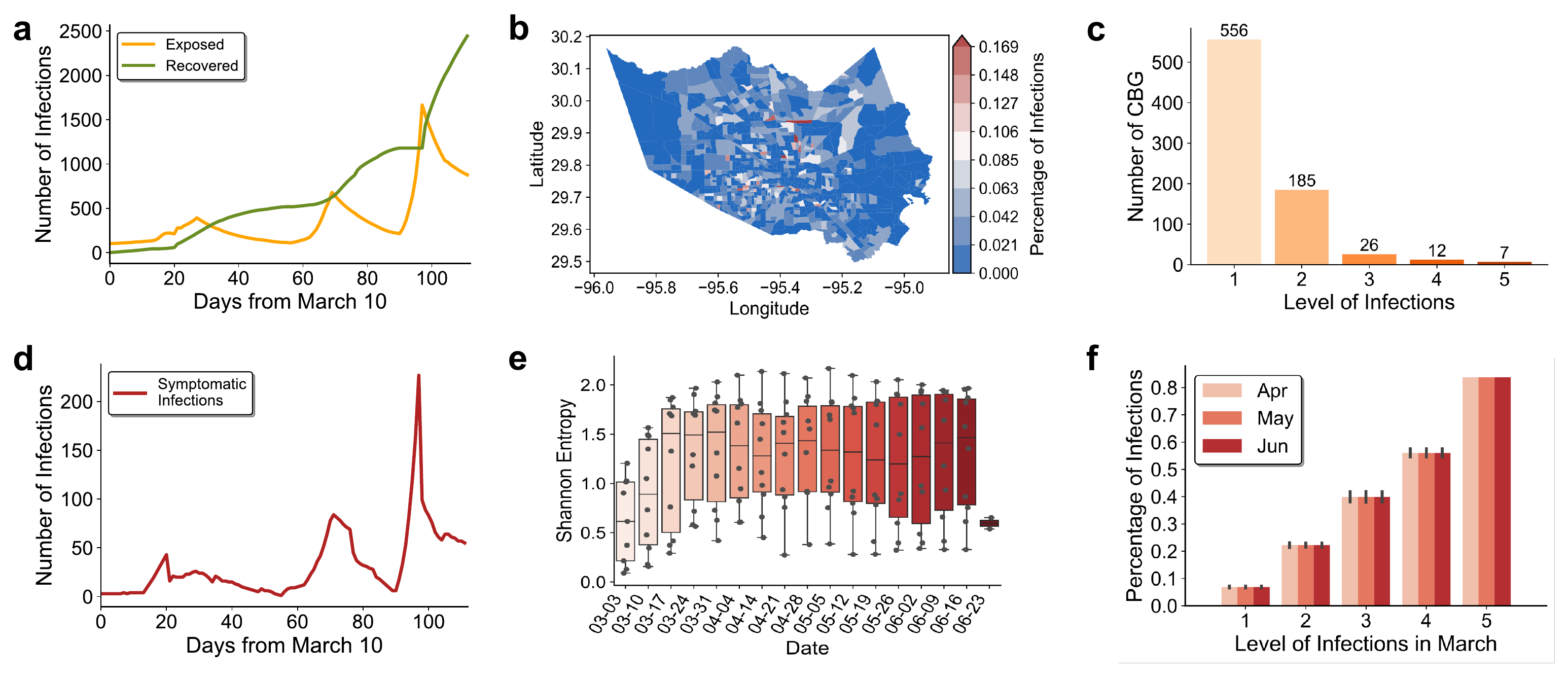}
  \caption{The outcomes of disease spread in the data-driven contact networks. {\sffamily \textbf{a.}} Daily number of exposed infections and cumulative number of recovered infections in Harris County, Texas, simulated by our model. {\sffamily \textbf{b.}} Spatial distribution of infections characterized by the percentage of people who were infected on the 80th day since March 10 across CBGs in Harris County, Texas, simulated by our model. Spatial distributions of infections in other counties are provided in supplementary information. {\sffamily \textbf{c.}} We categorize the CBGs in Harris County into five levels of infections. The range of percentages of infections at each level is the same. {\sffamily \textbf{d.}} Daily new symptomatic infections in Harris County, Texas, simulated by our model. {\sffamily \textbf{e.}} The results of Shannon Entropy quantifying the spatial heterogeneity of infections across CBGs in each county for the ten studied counties. {\sffamily \textbf{f.}} The percentage of infections in April, May, and June in CBGs at levels of infections determined in March.}
  \label{fig:fig2}
\end{figure}

We explain an example of the COVID-19 pandemic in Harris County (Houston area), Texas. Figures 3a and 3d show the predicted numbers of agents in exposed, removed, and symptomatic infectious states by fitting the model with reported cases. We observed that the pandemic situation evolved over time. For example, there are multiple peaks of infections in the study period. This result indicates that our model is distinct from existing simulation models and could capture realistic patterns of virus spread over time. Due to the lack of dynamic fined-grained contact networks, existing simulation models tend to simulate the spread of a disease on a static contact network in which the connections among the agents are not time-varying. Such simulations may not be able to account for the evolving human behaviors and explain the temporal variations and spatial heterogeneity of infections. 

Figure 3b shows an example of the geographical distribution of infections across Harris County. It is evident that the infections are geographically heterogeneous. This finding explains the contribution of contact patterns of people to the infection rates of their CBGs. People who engage in intensive contact activities cause their CBG to be intensively infected. The geographical heterogeneity in number of cases is further demonstrated in other counties (Figure 3e). The mean entropy of the infection distribution grows rapidly from 0.5 in the first week of March 2020 to around 1.5 in late March 2020, and it remains at a high level until the end of the study period. This result demonstrates that the infection patterns are extremely heterogeneous across different areas of the county and implementing uniform containment or reopening policies may lead to severe burdens on the economy or infection reduction efforts. Our model shows an important capability for uncovering the mechanism of disease spread, which could be used in formulating local policies. 

The results above raise an important need for localized policies, such as geo-fencing in high-infection CBGs and reopening in low-infection CBGs. Since policy development is based on the pandemic situation at the moment, the evaluation of policy effectiveness requires projection to the future. That is, the containment policies should apply to areas that show high infection at the current time step and will maintain a high level of infection in the future in the absence of containment policies. By doing so, the localized containment policies would have an optimal outcome in containing the spread of the disease. In response to this question, we split the range of infection percentages of the CBGs in a county into five levels with equal ranges. The example for Harris County is shown in in Figure 3c. We find that the numbers of cases across the CBGs are highly unequally distributed in these five levels. Most of the CBGs are at the lowest infection level, while a few CBGs are greatly infected. Then, we examine the level of infections of these CBGs in different months (Figure 3f). We observe little variances regarding the infection levels of the March CBGs in each group. This result demonstrates that the CBGs maintain their levels of infection during the pandemic, implying that adopting targeted policies for the CBGs with high infection levels at this time step is essential to preventing further deterioration of these CBGs. This finding provides important evidence for developing localized policies, which will be discussed in the following section.

\subsection*{\sffamily Effects of local policies on network structure and daily infections}
\vspace{-0.7em}
With the unveiled spatial-temporal patterns of the pandemic, local policies should account for such heterogeneous impacts. Since the outbreak began in early March 2020, governments and policymakers have introduced large-scale restriction measures, such as mobility control and mask use. Typical measures in mobility control include area lockdown and stay-at-home orders. The restriction of human contact activities, however, induced a heavy burden on the local economy, especially on labor-intensive businesses. To mitigate the economic pressures, many counties and states lifted their restrictions and allowed local businesses to reopen. These above-mentioned local policies were commonly adopted in the majority of counties in the United States. The local policies and their effects on contact networks lead to multiple peaks in the pandemic. This situation prompts a question regarding the extent to which these policies would reduce the pandemic spread. 

For these reasons, we project the direct impacts of typical local policies on human contact networks and estimate the consequences of contact network changes on infections. Three different scenarios through the implementations of three local policies are summarized here: 

\begin{itemize}
    \item \emph{Mobility control (CBG lockdown) scenario}: Mobility control measures including business lockdown and stay-at-home orders with the intent of limiting the number of contacts a person would have. In particular, local containment policies are developed to contain the activities of people from specific areas. Together with the findings from the last section, we rank the CBGs based on their percentage of infections from high to low. Then the mobility control measures are simulated by removing the connections of agents belonging to specific CBGs. We started containing the activities of agents from CBGs with a high level of infections, and selected 0\%, 25\%, 50\%, 75\% and 100\% of CBGs to be contained.
    \item \emph{Mask use scenario}: The SARS-CoV-2 infection is transmitted predominately by respiratory droplets generated when people talk to each other \cite{Moghadas2020,Worby2020}. The longer the face-to-face communication, the higher the chance of infection. Masks are primarily intended to reduce the emission of virus-laden droplets and reduce inhalation of these droplets by the wearer \cite{Fischer2020}. Hence, we simulate the mask use scenario by reducing the weights of the links between the agents with masks. If the weights after the reduction of a specific value become negative, we consider there is no link that would allow a disease to be transmitted. As such, with a certain percentage of agents wearing masks, the contact network would be sparser, implying a lower probability that a disease can transmit across the networks. It should be noted that the selection of agents wearing masks is random, without consideration of their CBGs and infection levels.
    \item \emph{Reopening scenario}: When reopening policies are enacted, businesses reopen and population activity increases. Reopening policies allow people to re-engage in contact activities to the level of normal conditions \cite{Vermund2020}. Hence, to simulate the scenario of reopening, we extracted the contact networks in February before the outbreak of the pandemic and synthesized connections similar to that of normal conditions for selected agents. The localized reopening policy is also executed in the CBGs in terms of their levels of infections. That is, we start reopening CBGs with the lowest level of infections then selecting CBGs with increasingly higher levels of infection based on specified reopening rate (percentage of reopened CBGs among all CBGs). In this study, the reopening rates are 0\%, 25\%, 50\%, 75\% and 100\%.
\end{itemize}

Figure 4 shows the effects of local policies on the structure of contact networks and the consequences of network changes on disease spread in Harris County. (See results for other counties in the supplementary information.) The first panel of the figure shows the effects of mobility control measures. As shown in Figure 4d, the total weights of the contact networks in Harris County decreased 50\% in about one month since March 10, 2020 and remained stable until the end of the study period. Mobility control measures implemented on the 13th week since March 10 significantly reduced the total weighting in the network. The reduction of the weighting is almost proportional to the percentage of the controlled CBGs in a county. When all CBGs are contained, the total weights of the network are 0, meaning that no links exist present in the contact network. This would be an ideal outcome of local containment, which cuts all possible transmission trajectories of the disease. Hence, mobility control policies could lead to promising results in terms of pandemic containment (Figure 4a). All curves with mobility control policies indicate a peak of infection one or two weeks after the policy implementation, as is shown in historical data. That is because the frequent contact activities enable a denser contact network, while the total weight of the network does not change dramatically. This conclusion is realistic as it indicates the effect of contact activities in disease spread. With the increase of controlled mobility in CBGs, the magnitude of the peak infections decreases. When 75\% or more CBGs are controlled for mobility, the peak number of infections is at a similar level as or even a lower level than the previous peak. This result implies that an aggressive mobility control policy is effective and necessary to change the trajectory of the pandemic, or at least to prevent the pandemic from worsening. 

The middle panel of Figure 4 shows the effects and consequences of mask use on the pandemic spread. When agents are in contact with others, masks reduce infection to a certain degree. Hence, the use of masks does not have evident effects (Figure 4d), as mobility control policies explicitly cut the connections among the agents and break down the contact network. Mask use is still essential since it can to some degree moderate the pandemic (Figure 4b). An increase in the number of people wearing masks leads to a significant decrease in the daily new infections and the magnitude of the peak number of infections. In an extreme case in which 90\% agents wear masks, the pandemic would be still out-of-control. That is because, there are some agents who contact others frequently and for a long time and some of them are infectious. The role of wearing masks is weakened in such cases. Hence, simply requiring all people to wear masks is insufficient to contain the epidemic.

The last panel of the figure indicates the outcomes of reopening policies. Although the stay-at-home order has been lifted since late May and June 2020, population mobility and contact activities did not revert to their pre-pandemic patterns. As we simulate the contact activities of some agents based on their regular activities in normal conditions, the contact networks get denser and the weights of the network increase drastically (Figure 4f). Such change in the network causes a sharp increase of the daily infections, two or three times the historical reported cases. In addition, as the hardest-hit CBGs reopen, the growth of daily infections rises even faster. This result shows a warning that untimely reopening policies during the pandemic could neutralize all previous control efforts. 

In summary, we have observed diverse effects of different policies and various levels of implementations of these policies on the structure of contact networks and the pandemic spread. Purely from a perspective of pandemic containment, both mobility control and mask use are necessary and effective in containing the transmission of the disease through population contact behaviors. During pandemic peak and economic recession, however, recovery of economic activities is also of great concern. Despite these exigencies, we must still take into account the larger economic loss which may be caused by a possible worsening pandemic in the future if we simply reopen regions without adequate containment measures \cite{Fan2021}.

\begin{figure}
  \centering
  \includegraphics[width=17cm]{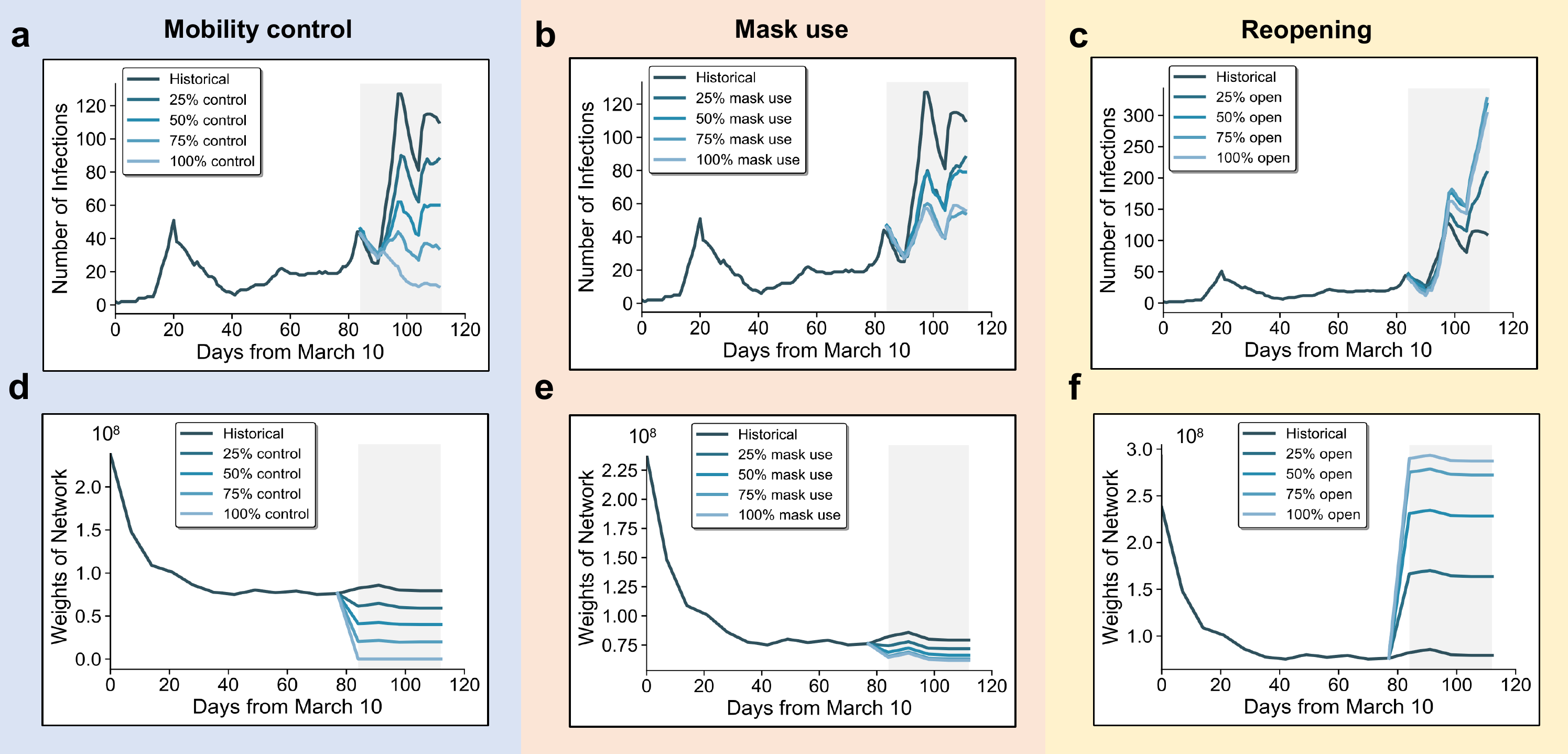}
  \caption{Effects of localized policies on daily infections and network structure for a variety of policy implementation levels. Each panel represents effectiveness in terms of relative levels of policy implementation on number of daily infections and the contact network. The impacts of varying policy levels on daily new infections {\sffamily \textbf{a-c}} are shown for different policies: mobility control, mask use, and reopening. The outcomes of varying policy levels on the weighted contact networks {\sffamily \textbf{d-f}} are shown with different policies. This study selects different levels of 0\%, 25\%, 50\%, 75\% and 100\% for policy implementation on the synthetic population.}
  \label{fig:fig3}
\end{figure}

\subsection*{\sffamily Impacts of combined policies on costs and infections}
\vspace{-0.7em}
The above analyses and results raise two important questions: what combination of policies would be more effective in containing the pandemic; and how we can make a trade-off between costs and infections for policy development. In this section, we present a method to quantify the costs of combined policies and also project the pandemic situation after implementing these combined policies. 

In our simulation model, we assume the system is consistent, meaning that no inflow nor outflow of agents occurs during the study period. The total number of agents remains the same throughout the study period. In addition, combined policies are consistently implemented during the test period. Thus, the timing of the policy implementation is not a factor influencing the calculation of the costs and the projection of infections. Based on these assumptions, we consider the costs of the policies to be proportional to the size of the population affected by the policies. Accordingly, we can assign a cost factor to each group of the population under specific policies to roughly quantify the costs of the policies. The cost function is formulated as follows:

\begin{equation}
\mathcal{C} = \mu_1 Pop_m + \mu_2 Pop_c - \mu_3 Pop_r + \mu_4 Pop_i
\end{equation}

where $\mu_1$, $\mu_2$ and $\mu_3$ are the cost factors for three types of populations under the policies. The values of these cost factors can be selected based on an estimation of the costs in specific local regions. Hence, the cost factors may be variant from region to region. The factors such as hospitalization, hospital capacity, age-dependent severity of illness and adherence with public health measures could also be incorporated in the cost factors to enable an accurate estimation. See more details in the supplementary information. In addition, $Pop_m$ is the percentage of the population wearing masks, $Pop_c$ is the percentage of the population in the CBGs with mobility control, $Pop_r$ is the percentage of the population in reopened CBGs, and $Pop_i$ is the number of new infections during the implementation of the policies. To compute the costs and project the infections for all combinations of the policies, we conducted a test for several levels of policy implementation. Specifically, we selected the values for three policy-related variables (i.e., $Pop_m$, $Pop_c$, and $Pop_r$) from 0.1 to 0.9 with increments of 0.1, estimated the infections under the specified policies, and computed the costs for all combinations of these variables. 

To compare costs of different local policies, we focused mainly on relative values of costs, which indicate extra savings or expenses of one policy compared to another. In other words, the cost values are not required to be precise in this analysis since relative costs are sufficient for us to identify cost-effective policies. As long as relative costs are reasonable, the assumption is defensible. Hence, we define $\mu_1$ to be 5, $\mu_2$ to be 30, $\mu_3$ to be 5, and $\mu_4$ to be 1. Taking the example of Harris County, Texas, it is clear that different combinations of local policies result in varying levels of infections in the test period (Figure 5a). Furthermore, by maintaining the stability of one policy, we can observe the cost and infection changes brought by the changes in the other two policies. For example, as we keep 10\% of reopening in CBGs at the lowest level of infection, a low rate of mobility control and mask use leads to a great number of infections (Figure 5f). The number of infections decreases as we increase mobility control and mask use. Meanwhile, the containment effect of mobility control is more evident than mask use. Such a pattern of effects by local policies influences the costs. As shown in Figure 5b, the most cost-effective strategy in this case is to implement mobility control (i.e., geo-fencing) in 10\% of CBGs which are highly infected with an 80\% mask-wearing rate. This result highlights the importance of mask use to achieve pandemic containment. In another case, we consider the rate of mobility control to be stable (10\%) and change the rates of CBGs with reopening and the percentage of people wearing masks. We find that the reopening of CBGs significantly worsens the epidemic situation, leading to an explosion of infections (Figure 5g). Due to the reopening and associated contact activities, the actual cost of a high rate of reopening is quite low, although a large number of people get infected (Figure 5c). An optimal strategy would be increasing the percentage of the population wearing masks and reopen the majority of the CBGs that are not highly infected. Finally, we examine the impact of changing the rates of mobility control and reopening and maintaining 10\% of the population wearing masks. Figure 5h shows a clear pattern that the upper left corner is dark, while the lower right corner is bright, indicating that reopening establishes a number of connections between people and subsequently provides a path for disease spread. Due to the effect of reopening on economic recovery, the cost could be relatively low when 10\% of the hardest-hit CBGs are under mobility control (Figure 5d).

Counties have different pandemic situations, which require different compound policies for containment. In this step, we apply our method to ten counties to identify their cost-effective combined policies. The optimal combined policy strategies are plotted in Figure 5e. The cost-effective combined policies for the majority of the counties, such as Dallas County, Texas (FIPS county code 48113, Dallas) and Wayne County, Michigan (FIPS county code 26163, Detroit) need a low percentage of CBG reopening, a high proportion of population wearing masks, and a large number of CBGs under mobility control. These counties tend to be among the hardest hit during the test period. Other counties, such as King County, Washington (53033, Seattle), and Suffolk County, Massachusetts (25025, Boston), require reopening for fewer infected CBGs to recover their economy, but severely infected CBGs should remain under mobility control. These findings indicate that mobility control is not always a fit-to-all option and not always cost-effective. Different counties require specific combined policies for containing the pandemic and to efficiently recover economic activities.

\begin{figure}
  \centering
  \includegraphics[width=17cm]{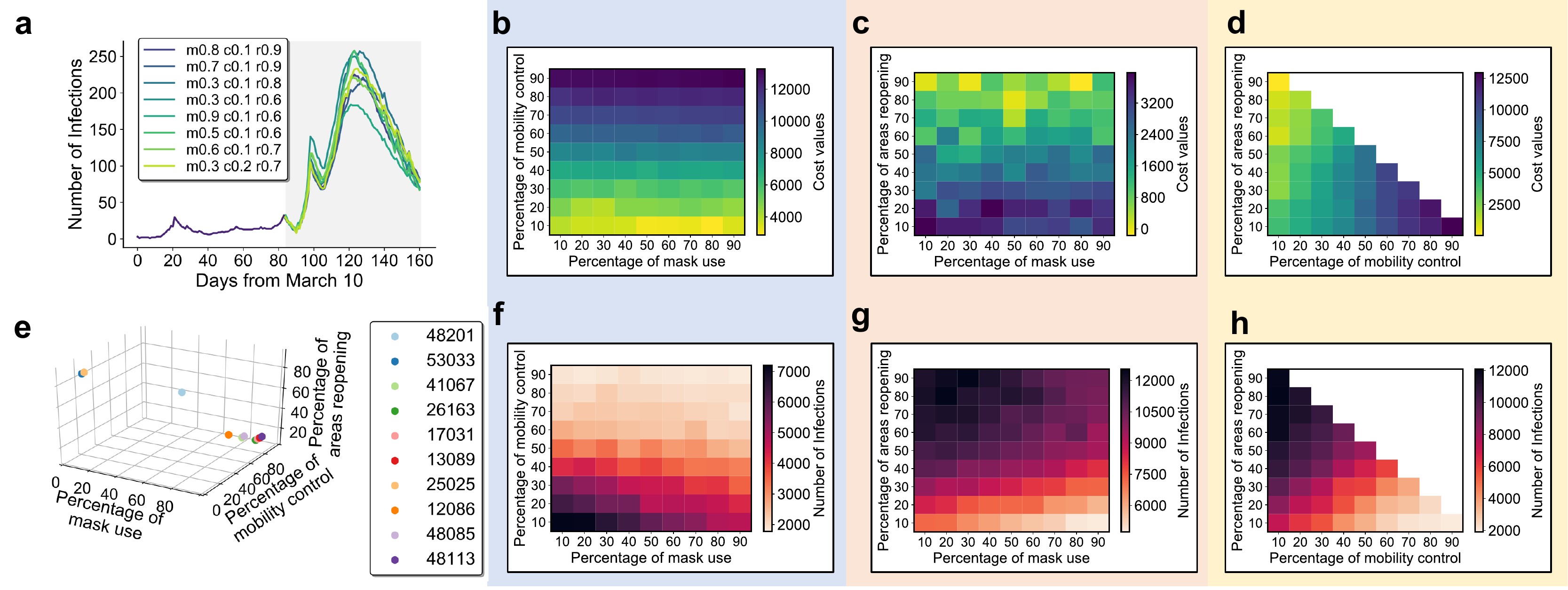}
  \caption{Outcomes of combined local containment policies and searching for optimal combinations. {\sffamily \textbf{a.}} Daily new infections under varying levels of compound policies (selected examples shown) in Harris County (FIPS county code 48201), Texas. To be visually clear, the plot shows only the outcomes for eight combinations of local policies. The estimated cost {\sffamily \textbf{b-d}} and the total number of infections {\sffamily \textbf{f-h}} with varying levels of compound policies. By keeping the reopen rate at 10\%, the cost and rate infections for varying levels of mobility control and mask use are shown in {\sffamily \textbf{b}} and {\sffamily \textbf{f}}, respectively. By keeping the mobility control rate to be 10\%, the cost and infection rate for varying levels of reopening and mask use are shown in {\sffamily \textbf{c}} and {\sffamily \textbf{g}}, respectively. By keeping the rate of mask use to be 10\%, the cost and infection rate for varying levels of reopening and mobility control are shown in {\sffamily \textbf{d}} and {\sffamily \textbf{h}}. {\sffamily \textbf{e}}, The optimal combination of the three local policies for the selected ten counties.}
  \label{fig:fig4}
\end{figure}

\section*{\sffamily Discussion and Concluding Remarks}
\vspace{-1em}
We present a data-driven contact network model that incorporates an epidemic contagion process. Using the synthetic contact network built based on anonymized data related to the contact activities from millions of people in ten major infected counties during the COVID-19 pandemic, our model performs well in predicting the spread of the disease in contact networks. Because our simulation is agent-based, we can identify the CBGs of the infected agents and assess the spatial-temporal heterogeneity of the pandemic in a county. This outcome is of particular importance since it addresses the challenge of geographical spread of the disease. As the model suggests, cross-CBG contact activities intensify disease transmission and enable a prediction of the infection distribution, which is consistent with the results of existing studies at national \cite{Jia2020} and international \cite{Wells2020} scales. The findings regarding the spatial-temporal patterns of the pandemic suggest localized containment strategies and inform us in which fraction of the population of urban areas the policies should be implemented. Our study can inform policymakers seeking to adopt area-specific policies instead of county-wide uniform measures \cite{chen2020bayesian}.

With the prevalence of mobility control, mask use and reopening policies \cite{Schunemann2020}, we examined the impacts of these policies on the structure of contact networks and the consequences of structural changes on epidemic spread. We find that mobility control is one of the most effective measures in containing the pandemic, yet imposes a heavy burden on the economy and causes onerous societal disruptions. This conundrum raises an urgent need for combined policies to balance the trade-off between economic aspects and infection cases. Our results show that the cost-effective strategy varies across different counties based on the local epidemic situation. Large-scale mobility control is not fit-to-all. An appropriate composition of the policies may achieve a relatively low infection level and also a low cost to society in the case of partial reopening and economic recovery. Our results and findings could be informative for government agencies for local containment strategies and for public response planning. For example, policymakers may increase testing and quarantine in highly infected CBGs, while lifting control for less infected CBGs.

Our study also confronts some limitations, specifically in the data set we use. It should be noted that the data does not cover the contact activities of all populations. For example, the activities from people who do not opt-in to the data-sharing contract of the data provider cannot be captured. Contact activities that occurred in non-points of interest may not be included in building the contact links in the network. These limitations in the dataset notwithstanding, fine-grained mobile phone data is widely adopted in modeling epidemics and informing public health policies \cite{Oliver2020}. In addition, although our model can accurately predict the pandemic using contact networks, the results do not imply the exact causal relation between contact activities and epidemic spread. Further studies are needed in improving the model to infer the in-depth mechanisms of epidemic spread and to substantiate effective policies in pandemic containment.

\section*{\sffamily Methods}
\vspace{-0.5em}
\subsection*{\sffamily Data sets}
\vspace{-0.7em}
We use geolocation data provided by Veraset, Inc. \cite{Veraset2020}, a data company that collects anonymized location data from numerous applications and from a large number of devices. Veraset, Inc. collects 2 to 3 billion location event data points every day from more than 30M unique devices across the United States. The anonymized devices opted-in to provide access to their location data through thousands of apps and software development kits (SDKs) with Veraset. The data was shared under a strict contract with Veraset through their collaborative program in which they provide access to de-identified and privacy-enhanced mobility data for academic research only. In this study, we filter the data for ten of the most infected counties from January 1 through June 27, 2020. The data contains anonymized device IDs, timestamps, and precise geographical coordinates of dwelling points. We legally performed the analysis and were required never to attempt to identify any individuals. 

To identify the contact activities at points of interest, we employed the geolocations of the POIs from SafeGraph \cite{SafeGraph2020}, a data company that has documented the geographical information and business information about physical places in the United States. 

Demographic data at the census block group level are adopted from the US Census Bureau’s American Community Survey (ACS) \cite{UnitedStatesCensusBureau2019}. The population size of each CBG which we use to synthesize agents in our model is the most recent one-year estimate (2018). 

We calibrated our model using cumulative reported cases from the data dataset published by the New York Times \cite{TheNewYorkTimes2020}. The dataset documents cumulative COVID-19 infections every day for all the counties in the United States. We fit out the model on these cumulative daily data to acquire the values of model parameters.

\subsection*{\sffamily Data-driven contact network}
\vspace{-0.7em}
The model considers a weighted network $\mathcal{G}=(\mathcal{V},\mathcal{E}_t,\mathcal{W}_t)$ with stable agents $\mathcal{V}$, time-varying links $\mathcal{E}_t$, and link weights $\mathcal{W}_t$. We define the agents $\mathcal{V}$ and assigned them to specific CBGs based on the population size of each CBG documented in the ACS demographic data. Then using the dwelling locations of the devices provided in the Veraset data, we considered two devices to be in contact if they presented in the same POI at the same time. Here, the dwelling locations of the devices are defined according to the time the devices spent in a location. We use 5 mins as the threshold amount of time to filter out the locations where the devices might be waiting for traffic lights. The length of the overlap of their time spent in the POI is denoted as the duration of their contact. As such, we create an empirical weighted contact network among the devices in the Veraset data. In addition, we estimated the home CBGs of the devices. The links between any pair of CBGs can be extracted. Based on the proportion between the population size of empirical data and the synthetic population, we have a scalar to quantify the difference between the number of devices and synthetic agents. The scalar is then applied to determine the number of links that should be synthesized for each pair of CBGs. The weights of these links follow the distribution of the weights in the empirical contact network. Next, we assign the links to the synthetic agents in these two CBGs, where the degree distribution for the agents is consistent with the degree distribution of the devices in the same CBGs in the empirical data. We generate and assign links by repeating the above steps for all pairs of CBGs. This approach allows for creating synthetic contact networks based on actual mobility data and to maintain key structural and attribute information of the agents and the networks.

We further conducted a test using Kullback-Leibler Divergence to ensure the generated contact networks remained the structural properties of the corresponding empirical network. The detailed results suggest that the degree distributions of the generated contact networks are similar to the degree distributions of the empirical networks (supplementary information). That is, our data-driven contact network is reliable for modeling realistic human contact patterns and pandemic spread.

\section*{\sffamily Data availability}
\vspace{-0.5em}
The data that support the findings of this study are available from Veraset Inc., but restrictions apply to the availability of these data, which were used under license for the current study. The data can be accessed upon request submitted on veraset.com. Other data we use in this study are all publicly available.

\section*{\sffamily Code availability}
\vspace{-0.5em}
The code that supports the findings of this study is available from the corresponding author upon request.

\renewcommand{\refname}{\large \sffamily References}
\bibliographystyle{unsrt}  
\bibliography{references}  

\begin{thebibliography}{10}

\bibitem{Headey2020}
Derek Headey, Rebecca Heidkamp, Saskia Osendarp, Marie Ruel, Nick Scott, Robert
  Black, Meera Shekar, Howarth Bouis, Augustin Flory, Lawrence Haddad, and Neff
  Walker.
\newblock {Impacts of COVID-19 on childhood malnutrition and nutrition-related
  mortality}.
\newblock {\em The Lancet}, 396(10250):519--521, aug 2020.

\bibitem{JohnHopkinsUniversity2020}
{John Hopkins University}.
\newblock {CSSEGISandData/COVID-19: Novel Coronavirus (COVID-19) Cases}, 2020.

\bibitem{Kraemer2020}
Moritz U~G Kraemer, Chia-Hung Yang, Bernardo Gutierrez, Chieh-Hsi Wu, Brennan
  Klein, David~M Pigott, Louis du~Plessis, Nuno~R Faria, Ruoran Li, William~P
  Hanage, John~S Brownstein, Maylis Layan, Alessandro Vespignani, Huaiyu Tian,
  Christopher Dye, Oliver~G Pybus, and Samuel~V Scarpino.
\newblock {The effect of human mobility and control measures on the COVID-19
  epidemic in China}.
\newblock {\em Science}, 368(6490):493 LP -- 497, may 2020.

\bibitem{Chinazzi2020}
Matteo Chinazzi, Jessica~T Davis, Marco Ajelli, Corrado Gioannini, Maria
  Litvinova, Stefano Merler, Ana {Pastore y Piontti}, Kunpeng Mu, Luca Rossi,
  Kaiyuan Sun, C{\'{e}}cile Viboud, Xinyue Xiong, Hongjie Yu, M~Elizabeth
  Halloran, Ira~M Longini, and Alessandro Vespignani.
\newblock {The effect of travel restrictions on the spread of the 2019 novel
  coronavirus (COVID-19) outbreak}.
\newblock {\em Science}, 368(6489):395 LP -- 400, apr 2020.

\bibitem{Rader2020}
Benjamin Rader, Samuel~V Scarpino, Anjalika Nande, Alison~L Hill, Ben Adlam,
  Robert~C Reiner, David~M Pigott, Bernardo Gutierrez, Alexander~E Zarebski,
  Munik Shrestha, John~S Brownstein, Marcia~C Castro, Christopher Dye, Huaiyu
  Tian, Oliver~G Pybus, and Moritz U~G Kraemer.
\newblock {Crowding and the shape of COVID-19 epidemics}.
\newblock {\em Nature Medicine}, 2020.

\bibitem{Polyakova2020}
Maria Polyakova, Geoffrey Kocks, Victoria Udalova, and Amy Finkelstein.
\newblock {Initial economic damage from the COVID-19 pandemic in the United
  States is more widespread across ages and geographies than initial mortality
  impacts}.
\newblock {\em Proceedings of the National Academy of Sciences}, page
  202014279, oct 2020.

\bibitem{AURAY2020104260}
St{\'{e}}phane Auray and Aur{\'{e}}lien Eyquem.
\newblock {The macroeconomic effects of lockdown policies}.
\newblock {\em Journal of Public Economics}, 190:104260, 2020.

\bibitem{Jia2020}
Jayson~S Jia, Xin Lu, Yun Yuan, Ge~Xu, Jianmin Jia, and Nicholas~A Christakis.
\newblock {Population flow drives spatio-temporal distribution of COVID-19 in
  China}.
\newblock {\em Nature}, 2020.

\bibitem{Balcan2009}
Duygu Balcan, Vittoria Colizza, Bruno Gon{\c{c}}alves, Hao Hu, Jos{\'{e}}~J
  Ramasco, and Alessandro Vespignani.
\newblock {Multiscale mobility networks and the spatial spreading of infectious
  diseases}.
\newblock {\em Proceedings of the National Academy of Sciences}, 106(51):21484
  LP -- 21489, dec 2009.

\bibitem{Newman2002}
M.~E.J. Newman.
\newblock {Spread of epidemic disease on networks}.
\newblock {\em Physical Review E}, 2002.

\bibitem{Chowell2004}
G~Chowell, N~W Hengartner, C~Castillo-Chavez, P~W Fenimore, and J~M Hyman.
\newblock {The basic reproductive number of Ebola and the effects of public
  health measures: the cases of Congo and Uganda}.
\newblock {\em Journal of Theoretical Biology}, 229(1):119--126, 2004.

\bibitem{Ramchandani2020}
A~Ramchandani, C~Fan, and A~Mostafavi.
\newblock {DeepCOVIDNet: An Interpretable Deep Learning Model for Predictive
  Surveillance of COVID-19 Using Heterogeneous Features and Their
  Interactions}.
\newblock {\em IEEE Access}, 8:159915--159930, 2020.

\bibitem{kapoor2020examining}
Amol Kapoor, Xue Ben, Luyang Liu, Bryan Perozzi, Matt Barnes, Martin Blais, and
  Shawn O'Banion.
\newblock {Examining COVID-19 Forecasting using Spatio-Temporal Graph Neural
  Networks}, 2020.

\bibitem{Chang2020}
Serina Chang, Emma Pierson, Pang~Wei Koh, Jaline Gerardin, Beth Redbird, David
  Grusky, and Jure Leskovec.
\newblock {Mobility network models of COVID-19 explain inequities and inform
  reopening}.
\newblock {\em Nature}, 2020.

\bibitem{Holtz2020}
David Holtz, Michael Zhao, Seth~G Benzell, Cathy~Y Cao, Mohammad~Amin Rahimian,
  Jeremy Yang, Jennifer Allen, Avinash Collis, Alex Moehring, Tara Sowrirajan,
  Dipayan Ghosh, Yunhao Zhang, Paramveer~S Dhillon, Christos Nicolaides, Dean
  Eckles, and Sinan Aral.
\newblock {Interdependence and the cost of uncoordinated responses to
  COVID-19}.
\newblock {\em Proceedings of the National Academy of Sciences}, 117(33):19837
  LP -- 19843, aug 2020.

\bibitem{schlosser2020covid19}
Frank Schlosser, Benjamin~F Maier, Olivia Jack, David Hinrichs, Adrian
  Zachariae, and Dirk Brockmann.
\newblock {COVID-19 lockdown induces disease-mitigating structural changes in
  mobility networks}, 2020.

\bibitem{Oliver2020}
Nuria Oliver, Bruno Lepri, Harald Sterly, Renaud Lambiotte, S{\'{e}}bastien
  Delataille, Marco {De Nadai}, Emmanuel Letouz{\'{e}}, Albert~Ali Salah,
  Richard Benjamins, Ciro Cattuto, Vittoria Colizza, Nicolas de~Cordes,
  Samuel~P Fraiberger, Till Koebe, Sune Lehmann, Juan Murillo, Alex Pentland,
  Phuong~N Pham, Fr{\'{e}}d{\'{e}}ric Pivetta, Jari Saram{\"{a}}ki, Samuel~V
  Scarpino, Michele Tizzoni, Stefaan Verhulst, and Patrick Vinck.
\newblock {Mobile phone data for informing public health actions across the
  COVID-19 pandemic life cycle}.
\newblock {\em Science Advances}, page eabc0764, apr 2020.

\bibitem{Grantz2020}
Kyra~H Grantz, Hannah~R Meredith, Derek A~T Cummings, C~Jessica~E Metcalf,
  Bryan~T Grenfell, John~R Giles, Shruti Mehta, Sunil Solomon, Alain Labrique,
  Nishant Kishore, Caroline~O Buckee, and Amy Wesolowski.
\newblock {The use of mobile phone data to inform analysis of COVID-19 pandemic
  epidemiology}.
\newblock {\em Nature Communications}, 11(1):4961, 2020.

\bibitem{Liu2018a}
Quan-Hui Liu, Marco Ajelli, Alberto Aleta, Stefano Merler, Yamir Moreno, and
  Alessandro Vespignani.
\newblock {Measurability of the epidemic reproduction number in data-driven
  contact networks}.
\newblock {\em Proceedings of the National Academy of Sciences}, 115(50):12680
  LP -- 12685, dec 2018.

\bibitem{Aleta2020a}
Alberto Aleta, David Mart{\'{i}}n-Corral, Ana {Pastore y Piontti}, Marco
  Ajelli, Maria Litvinova, Matteo Chinazzi, Natalie~E Dean, M~Elizabeth
  Halloran, Ira~M {Longini Jr}, Stefano Merler, Alex Pentland, Alessandro
  Vespignani, Esteban Moro, and Yamir Moreno.
\newblock {Modelling the impact of testing, contact tracing and household
  quarantine on second waves of COVID-19}.
\newblock {\em Nature Human Behaviour}, 4(9):964--971, 2020.

\bibitem{Bonaccorsi2020}
Giovanni Bonaccorsi, Francesco Pierri, Matteo Cinelli, Andrea Flori, Alessandro
  Galeazzi, Francesco Porcelli, Ana~Lucia Schmidt, Carlo~Michele Valensise,
  Antonio Scala, Walter Quattrociocchi, and Fabio Pammolli.
\newblock {Economic and social consequences of human mobility restrictions
  under COVID-19}.
\newblock {\em Proceedings of the National Academy of Sciences}, 117(27):15530
  LP -- 15535, jul 2020.

\bibitem{Kaxiras2020}
Efthimios Kaxiras and Georgios Neofotistos.
\newblock {Multiple Epidemic Wave Model of the COVID-19 Pandemic: Modeling
  Study}.
\newblock {\em Journal of medical Internet research}, 22(7):e20912--e20912, jul
  2020.

\bibitem{Bureau}
US~Census Bureau.
\newblock {2019 National and State Population Estimates}.

\bibitem{CentersforDiseaseControlandPrevention2020}
{Centers for Disease Control and Prevention}.
\newblock {Scientific Brief: SARS-CoV-2 and Potential Airborne Transmission},
  oct 2020.

\bibitem{PhysRevE.63.066117}
Romualdo Pastor-Satorras and Alessandro Vespignani.
\newblock {Epidemic dynamics and endemic states in complex networks}.
\newblock {\em Phys. Rev. E}, 63(6):66117, may 2001.

\bibitem{barabasi2016network}
Albert-L{\'{a}}szl{\'{o}} Barab{\'{a}}si and M{\'{a}}rton P{\'{o}}sfai.
\newblock {\em {Network science}}.
\newblock Cambridge University Press, Cambridge, 2016.

\bibitem{Guan2020}
Wei-jie Guan, Zheng-yi Ni, Yu~Hu, Wen-hua Liang, Chun-quan Ou, Jian-xing He,
  Lei Liu, Hong Shan, Chun-liang Lei, David~S.C. Hui, Bin Du, Lan-juan Li,
  Guang Zeng, Kwok-Yung Yuen, Ru-chong Chen, Chun-li Tang, Tao Wang, Ping-yan
  Chen, Jie Xiang, Shi-yue Li, Jin-lin Wang, Zi-jing Liang, Yi-xiang Peng,
  Li~Wei, Yong Liu, Ya-hua Hu, Peng Peng, Jian-ming Wang, Ji-yang Liu, Zhong
  Chen, Gang Li, Zhi-jian Zheng, Shao-qin Qiu, Jie Luo, Chang-jiang Ye,
  Shao-yong Zhu, and Nan-shan Zhong.
\newblock {Clinical Characteristics of Coronavirus Disease 2019 in China}.
\newblock {\em New England Journal of Medicine}, 382(18):1708--1720, apr 2020.

\bibitem{Aleta2020}
Alberto Aleta, David Martin-Corral, Ana {Pastore y Piontti}, Marco Ajelli,
  Maria Litvinova, Matteo Chinazzi, Natalie~E Dean, M~Elizabeth Halloran, Ira~M
  Longini, Stefano Merler, Alex Pentland, Alessandro Vespignani, Esteban Moro,
  and Yamir Moreno.
\newblock {Modeling the impact of social distancing, testing, contact tracing
  and household quarantine on second-wave scenarios of the COVID-19 epidemic}.
\newblock {\em medRxiv}, page 2020.05.06.20092841, jan 2020.

\bibitem{Healthcareworkers}
Healthcare workers.
\newblock {Duration of Isolation and Precautions for Adults with COVID-19}.

\bibitem{TheNewYorkTimes2020}
{The New York Times}.
\newblock {Covid in the U.S.: Latest Map and Case Count - The New York Times},
  2020.

\bibitem{Moghadas2020}
Seyed~M. Moghadas, Meagan~C. Fitzpatrick, Pratha Sah, Abhishek Pandey, Affan
  Shoukat, Burton~H. Singer, and Alison~P. Galvani.
\newblock {The implications of silent transmission for the control of COVID-19
  outbreaks}.
\newblock {\em Proceedings of the National Academy of Sciences of the United
  States of America}, 117(30):17513--17515, jul 2020.

\bibitem{Worby2020}
Colin~J Worby and Hsiao-Han Chang.
\newblock {Face mask use in the general population and optimal resource
  allocation during the COVID-19 pandemic}.
\newblock {\em Nature Communications}, 11(1):4049, 2020.

\bibitem{Fischer2020}
Emma~P. Fischer, Martin~C. Fischer, David Grass, Isaac Henrion, Warren~S.
  Warren, and Eric Westman.
\newblock {Low-cost measurement of face mask efficacy for filtering expelled
  droplets during speech}.
\newblock {\em Science Advances}, 6(36), sep 2020.

\bibitem{Vermund2020}
Sten~H Vermund and Virginia~E Pitzer.
\newblock {Asymptomatic Transmission and the Infection Fatality Risk for
  COVID-19: Implications for School Reopening}.
\newblock {\em Clinical Infectious Diseases}, jun 2020.

\bibitem{Fan2021}
Chao Fan, Sanghyeon Lee, Yang Yang, Bora Oztekin, Qingchun Li, and Ali
  Mostafavi.
\newblock {Effects of population co-location reduction on cross-county
  transmission risk of COVID-19 in the United States}.
\newblock {\em Applied Network Science}, 6(1):14, 2021.

\bibitem{Wells2020}
Chad~R Wells, Pratha Sah, Seyed~M Moghadas, Abhishek Pandey, Affan Shoukat,
  Yaning Wang, Zheng Wang, Lauren~A Meyers, Burton~H Singer, and Alison~P
  Galvani.
\newblock {Impact of international travel and border control measures on the
  global spread of the novel 2019 coronavirus outbreak}.
\newblock {\em Proceedings of the National Academy of Sciences}, 117(13):7504
  LP -- 7509, mar 2020.

\bibitem{chen2020bayesian}
Peng Chen, Keyi Wu, and Omar Ghattas.
\newblock {Bayesian inference of heterogeneous epidemic models: Application to
  COVID-19 spread accounting for long-term care facilities}, 2020.

\bibitem{Schunemann2020}
Holger~J Sch{\"{u}}nemann, Elie~A Akl, Roger Chou, Derek~K Chu, Mark Loeb,
  Tamara Lotfi, Reem~A Mustafa, Ignacio Neumann, Lynora Saxinger, Shahnaz
  Sultan, and Dominik Mertz.
\newblock {Use of facemasks during the COVID-19 pandemic}.
\newblock {\em The Lancet Respiratory Medicine}, 8(10):954--955, oct 2020.

\bibitem{Veraset2020}
Veraset.
\newblock {Veraset - Analyze + Predict + Build With Veraset's Leading Global
  Population Movement Data.}, 2020.

\bibitem{SafeGraph2020}
SafeGraph.
\newblock {SafeGraph: Places Data {\&} Foot-Traffic Insights}, 2020.

\bibitem{UnitedStatesCensusBureau2019}
{United States Census Bureau}.
\newblock {American Community Survey 2014-2018 5-Year Estimates}, dec 2019.

\end{thebibliography}

\section*{\sffamily Acknowledgements}
\vspace{-0.5em}
This material is based in part upon work supported by the National Science Foundation under Grant SES-2026814 (RAPID), the National Academies’ Gulf Research Program Early-Career Research Fellowship, the Amazon Web Services (AWS) Machine Learning Award, and the Microsoft AI for Public Health Grant. The authors also would like to acknowledge the data support from Veraset Inc. Any opinions, findings, conclusions or recommendations expressed in this material are those of the authors and do not necessarily reflect the views of the National Science Foundation, Amazon Web Services, Microsoft, SafeGraph or Veraset, Inc.


\section*{\sffamily Competing interests}
\vspace{-0.5em}
The authors declare that they have no competing interests.






\end{document}